\documentclass[notitlepage,12pt,tightenlines,eqsecnum,nofootinbib,floatfix,prd,letterpaper,superscriptaddress]{revtex4-1}

\usepackage{enumitem}
\usepackage{epsfig}
\usepackage{amsfonts}
\usepackage{amssymb}
\usepackage{amsbsy} 
\usepackage{graphicx}
\usepackage[usenames,dvipsnames]{color}
\usepackage{lastpage}
\usepackage{fancyhdr}
\usepackage{amsmath}
\usepackage{amsthm}
\usepackage{bm}
\usepackage{dcolumn}
\usepackage{epsfig}
\usepackage{graphics}
\usepackage[utf8]{inputenc}
\usepackage{latexsym}
\usepackage{hyperref}
\usepackage{tabularx} 
\usepackage{mathptmx}
\usepackage[margin=1in]{geometry}
\usepackage{setspace}
\usepackage{setspace}

\usepackage[table]{xcolor}
\usepackage{colortbl}
\usepackage{multirow}
\usepackage{ulem}

\newcommand{\be}{\begin{equation}}
\newcommand{\ee}{\end{equation}}
\newcommand{\ba}{\begin{eqnarray}}
\newcommand{\ea}{\end{eqnarray}}

\definecolor{darkred}{rgb}{0.8, 0.0, 0.0}
\definecolor{midred}{rgb}{0.94, 0.19, 0.22}
\definecolor{litered}{rgb}{0.91, 0.33, 0.5}
\definecolor{superlitered}{rgb}{0.99, 0.56, 0.67}
\definecolor{darkblue}{rgb}{0.0, 0.0, 0.55}
\definecolor{midblue}{rgb}{0.12, 0.56, 1.0}
\definecolor{liteblue}{rgb}{0.0, 0.72, 0.92}
\definecolor{superliteblue}{rgb}{0.49, 0.98, 1.0}
\definecolor{darkgreen}{rgb}{0.09, 0.45, 0.27}
\definecolor{midgreen}{rgb}{0.52, 0.73, 0.4}

\allowdisplaybreaks[4]

\AtBeginDocument{
  \addtocontents{toc}{\footnotesize}
  \addtocontents{lof}{\footnotesize}
}

\newcommand{\nocontentsline}[3]{}
\newcommand{\tocless}[2]{\bgroup\let\addcontentsline=\nocontentsline#1{#2}\egroup}

\begin{document}

\title{Astro2020 Science White Paper:\\Tests of General Relativity and Fundamental Physics with
  Space-based Gravitational Wave Detectors}

%\noindent
%\begin{center}
%Thematic Science Area: Cosmology and Fundamental Physics
%\end{center}
\def\checkbox{\square\!\!\!\!\!\!\!{\bf X}}

\noindent{\bf Thematic Areas:}
$$\vbox{\hsize 5truein \halign{# \hfil & #\hfil \cr
$\square$ Planetary Systems & $\square$ Star and Planet Formation \cr
$\square$ Formation and Evolution of Compact Objects & $\checkbox$ Cosmology and Fundamental Physics \cr
$\square$  Stars and Stellar Evolution & $\square$ Resolved Stellar Populations and their Environments \cr
$\square$ Galaxy Evolution & $\square$ Multi-Messenger Astronomy and Astrophysics \cr
}}$$

\begin{abstract}
  Low-frequency gravitational-wave astronomy can perform precision
  tests of general relativity and probe fundamental physics in a
  regime previously inaccessible. A space-based detector will be a
  formidable tool to explore gravity's role in the cosmos, potentially
  telling us if and where Einstein's theory fails and providing clues
  about some of the greatest mysteries in physics and astronomy, such
  as dark matter and the origin of the Universe.
\end{abstract}
\date{\today}

%\newpage

\author{Emanuele Berti}\email{email: berti@jhu.edu, phone: 410-516-2535}
\affiliation{Department of Physics and Astronomy, Johns Hopkins
  University, 3400 N. Charles Street, Baltimore, MD 21218, USA}
\affiliation{Department of Physics and Astronomy, The University of Mississippi, University, MS 38677, USA}

\author{Enrico Barausse}
\affiliation{CNRS, UMR 7095, Institut d’Astrophysique de Paris, 98 bis Bd Arago, 75014 Paris, France}
\affiliation{Sorbonne Universit\'es, UPMC Univesit\'e Paris 6, UMR 7095, Institut d’Astrophysique de Paris, 98 bis Bd Arago, 75014 Paris, France}

\author{Ilias Cholis}
\affiliation{Department of Physics, Oakland University, Rochester, MI 48309 USA}

\author{Juan Garc\'ia-Bellido}
\affiliation{Instituto de F\'isica Te\'orica UAM-CSIC, Universidad Auton\'oma de Madrid, Cantoblanco, 28049 Madrid, Spain}
\affiliation{CERN, Theoretical Physics Department, 1211 Geneva, Switzerland}

\author{Kelly Holley-Bockelmann}
\affiliation{Department of Physics and Astronomy, Vanderbilt University, 2301 Vanderbilt Place, Nashville, TN 37235, USA}
\affiliation{Department of Physics, Fisk University, 1000 17th Ave. N, Nashville, TN 37208, USA}

\author{Scott A. Hughes}
\affiliation{Department of Physics and MIT Kavli Institute, Cambridge, MA 02139}

\author{Bernard Kelly}
\affiliation{Gravitational Astrophysics Laboratory, NASA Goddard Space Flight Center, Greenbelt, MD 20771, USA}
\affiliation{CRESST, NASA Goddard Space Flight Center, Greenbelt, MD 20771, USA}
\affiliation{Department of Physics, University of Maryland, Baltimore County, Baltimore, MD 21250, USA}

\author{Ely D. Kovetz}
\affiliation{Department of Physics and Astronomy, Johns Hopkins University, 3400 N. Charles Street, Baltimore, MD 21218, USA}

\author{Tyson B. Littenberg}
\affiliation{NASA Marshall Space Flight Center, Huntsville AL 35812, USA}

\author{Jeffrey Livas}
\affiliation{Gravitational Astrophysics Laboratory, NASA Goddard Space Flight Center, Greenbelt, MD 20771, USA}

\author{Guido Mueller}
\affiliation{University of Florida, Gainesville, FL 32611, USA}

\author{Priya Natarajan}
\affiliation{Department of Astronomy, Yale University, 52 Hillhouse Avenue, New Haven, CT 06511}

\author{David H. Shoemaker}
\affiliation{LIGO, Massachusetts Institute of Technology, Cambridge, MA 02139, USA}

\author{Deirdre Shoemaker}
\affiliation{Center for Relativistic Astrophysics and School of Physics, Georgia Institute of Technology, Atlanta, GA 30332, USA}

\author{Jeremy D. Schnittman}
\affiliation{Gravitational Astrophysics Laboratory, NASA Goddard Space Flight Center, Greenbelt, MD 20771, USA}
\affiliation{Department of Physics, University of Maryland, Baltimore County, Baltimore, MD 21250, USA}

\author{Michele Vallisneri}
\affiliation{Jet Propulsion Laboratory, 4800 Oak Grove Drive, Pasadena, CA 91109, USA}

\author{Nicol\'as Yunes}
\affiliation{eXtreme Gravity Institute, Department of Physics, Montana State
University, Bozeman, Montana 59717, USA}

\maketitle

\newpage

%\maketitle
\thispagestyle{empty}

\pagestyle{fancy}
\rfoot{ \thepage }
\cfoot{}
\lhead{Berti et al.}
\rhead{Tests of General Relativity \& Fundamental Physics}

Einstein's theory of gravity, general relativity (GR), has been a
triumph of theoretical physics, having passed numerous observational
tests -- from the perihelion precession of Mercury's orbit around the
Sun, to the Nobel Prize winning discoveries of the Hulse-Taylor
pulsar~\cite{Hulse:1974eb} and of gravitational waves (GWs) from black
hole (BH) binary mergers~\cite{Abbott:2016blz}.  Nonetheless, there
are strong theoretical reasons -- which relate to the origin of the
Universe and physics beyond the Standard Model -- to suspect that a
deeper theory will emerge upon closer scrutiny.

Until recently, our picture of the Universe was mostly assembled using
traditional electromagnetic telescopes in every waveband, from gamma
rays to radio.  Although this picture revealed wonders, it has so far
lacked highly precise information about objects where gravity is
extremely strong such as BHs, or where gravity is dynamical and speeds
are relativistic.  The missing pieces of the puzzle will be provided
by GW detectors, such as LIGO~\cite{Abramovici:1992ah},
Virgo~\cite{Giazotto:1988gw}, and future ground-based detectors;
Pulsar Timing Arrays~\cite{Verbiest:2016vem}; and the planned future
space-based interferometers such as LISA~\cite{Audley:2017drz}.
Unlike light, GWs are not easily absorbed by matter, allowing us to
peer beyond interstellar gas, beyond intervening galaxies, beyond
accretion disks of massive BHs, and into the hearts of strong-gravity
objects.  This white paper concerns the theme \textit{Cosmology and
  Fundamental Physics}, and specifically addresses tests of gravity
and fundamental physics.  Another white paper, \textit{Cosmology with
  a space-based gravitational wave observatory} by Caldwell et al.,
focuses on the exciting opportunities provided by mHz GW observations
for understanding the early Universe.

Space-based GW observatories are sensitive to gravitational
wavelengths inaccessible to their ground-based
counterparts~\cite{Abramovici:1992ah}. Current detectors operate at
frequencies $\gtrsim 10$ Hz and detect binaries with masses
$\lesssim 10^{2} M_{\odot}$ in the ``local'' Universe. This includes
BH merger events lasting minutes or less, with typical signal-to-noise
ratio (SNRs) of a few tens. Space-based missions such as
LISA~\cite{AmaroSeoane:2012je,Audley:2017drz} will operate at
frequencies $\sim 1$ mHz and target very different source populations,
including the merger of massive BHs in galactic centers. Events can
last weeks, months or years with SNRs in the hundreds to thousands,
allowing us to probe a much larger volume of the Universe. Space-based
observatories will measure $\sim 10^4$--$10^5$ GW cycles from massive
BH mergers and extreme mass-ratio inspirals (EMRIs), encoding
information from which we will draw exquisitely precise astrophysical
measurements and perform stringent tests of GR in the strong gravity
regime.

{\bf The science case for space-based GW detectors as fundamental
  physics experiments is outstanding.}  Compared to binary pulsar and
ground-based observations, these detectors will provide high-SNR
observations of completely different source populations, potentially
revealing novel phenomena and probing gravity at very different
frequencies, and hence different source masses and energy scales. Some
GR modifications affect gravity only in the strong field, while
simultaneously passing all binary pulsar, cosmology and Solar System
tests~\cite{Will:2014kxa,Yunes:2013dva,Berti:2015itd}.  Space-based GW
detectors have an unprecedented potential to carry out
\textit{precision tests} in this mostly unexplored strong gravity
regime~\cite{Gair:2012nm}.  {\bf A single space-based detection should
  allow for precision tests of GR at the sub 0.1\% level, a factor of
  100--1000 better than current ground-based detections.}

Schematically, modifications of GR can affect: {\bf (a) GW generation,
  (b) GW propagation, (c) BH spacetimes,} and/or {\bf (d) BH
  dynamics}.  Modifications to GW generation can, for example, lead to
violations of the strong equivalence principle.  Modifications to the
propagation of GWs can be thought of as changes in their dispersion
relation, e.g.  due to gravitational Lorentz symmetry breaking.
Modifications to the astrophysical expectation that rotating BHs are
described by the Kerr metric arise in various gravity theories, such
as those containing high-order curvature terms in the action.
Finally, the relaxation of the remnant after a BH collision is encoded
in its oscillation spectrum (the so-called
``ringdown''~\cite{Berti:2009kk}), which is also typically corrected
in modified theories of gravity.  Different GW sources targeted by
space-based GW detectors are better at probing different classes of
modifications of GR. Massive BH mergers are excellent at probing (a),
(b), and (d). The inspiral of a small compact object into a massive BH
(a so-called extreme mass-ratio inspiral, or EMRI) will be very good
for (b) and has the unique ability to probe (c), as discussed below.

\begin{figure*}[h]
\vspace{-0.3cm}
\begin{center}
\begin{tabular}{cc}
\includegraphics[width=12cm,clip=true]{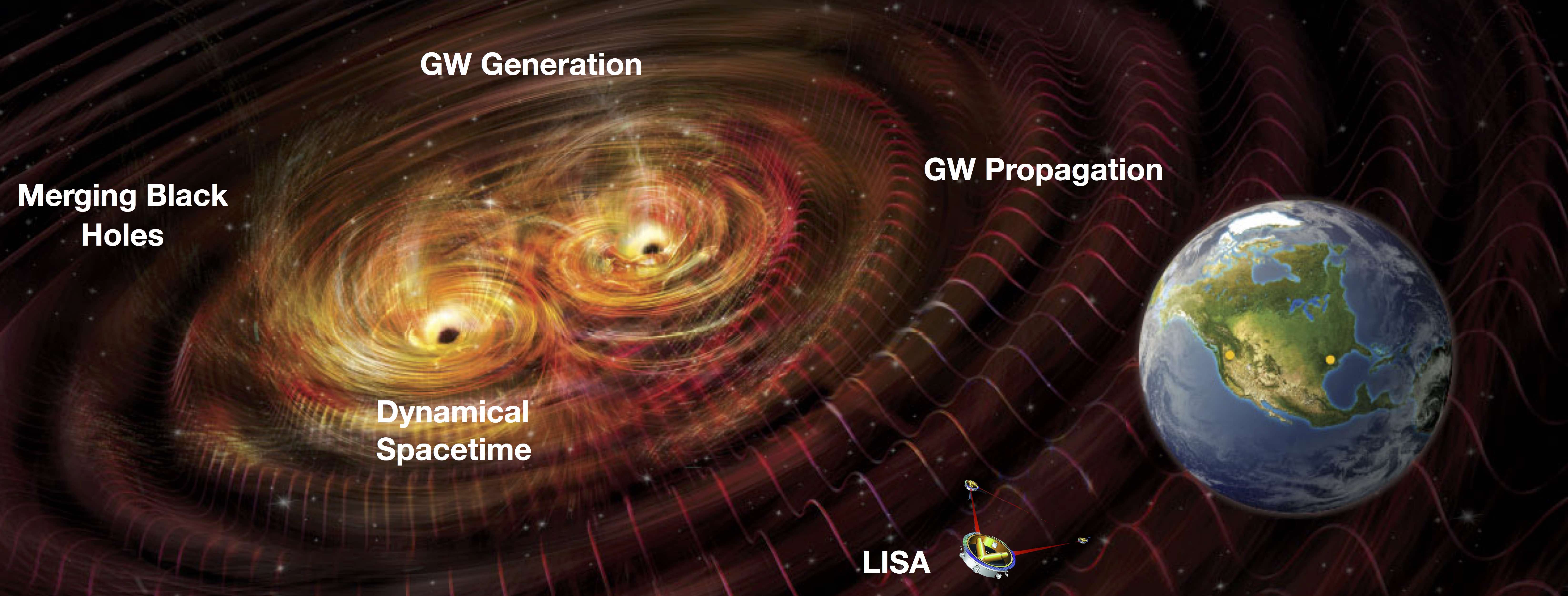} 
\end{tabular}
\end{center}
\vspace{-1cm}
\end{figure*}

%\vspace{.1cm}

\noindent
{\bf \textit{Probing gravitational wave generation and propagation.}}
Inferences drawn from GWs can be divided into two classes, depending
on the sector of the theory they constrain: \textit{(a) generation}
and \textit{(b) propagation}. The generation sector deals with the way
GWs (and any additional degree of freedom) are produced, and how they
evolve in time and backreact on the evolution of the source (say, a
binary system). The propagation sector deals with how GWs travel away
from the source.

It took the gravity theory community approximately 50 years to obtain
a model of the GWs emitted in the inspiral and merger of compact
binaries in GR accurate enough for LIGO/Virgo observations.  Given the
plethora of proposed modified gravity models and the extreme
difficulty in constructing sufficiently accurate GW models for data
analysis, it is not very useful to carry out similar calculations on a
theory-by-theory basis. It is much more appealing to develop generic
tests of Einstein’s theory given the available data. In the context of
Solar System observations one can expand the Einstein equations in
powers of $v/c$ around their Newtonian limit. This idea led to the
development of the widely used ``parameterized post-Newtonian'' (ppN)
framework of Will and
Nordtvedt~\cite{Nordtvedt:1968qs,1971ApJ...163..611W,1972ApJ...177..757W,1972ApJ...177..775N}.
A similar proposal to carry out generic tests was implemented in the
analysis of LIGO BH merger
observations~\cite{TheLIGOScientific:2016src}, and it consists of
verifying the post-Newtonian structure of the GW
phase~\cite{Arun:2006yw}. In the parameterized post-Einsteinian (ppE)
approach~\cite{Yunes:2009ke}, Bayesian inference on the data decides
the posteriors for the magnitude of ``GR correction'' parameters,
which can then be mapped to posteriors on coupling parameters of
specific theories~\cite{Yunes:2013dva,Berti:2018cxi}.

Such theory-agnostic approaches allow us to answer the question: {\bf
  what new physics can be probed with space-based GW detectors?}  Any
order-of-magnitude improvement in our understanding of the behavior of
strong, dynamical gravity can lead to potential breakthroughs. As an
illustration of the dangers of extrapolating physical theories, note
that the gravitational potential at the Earth’s surface (where
Newtonian gravity is extremely successful) is only 4 orders of
magnitude smaller than the gravitational potential at the Sun’s
surface, where relativistic effects are relevant (as shown by the
classical Solar System tests of GR). Going beyond gravity, a very
successful theory such as quantum electrodynamics cannot be
extrapolated from atomic to nuclear energy scales, where the strong
interaction dominates; again, these two scales are separated by just 6
orders of magnitude. Therefore it is not unreasonable to expect that
{\bf LISA may provide breakthroughs in our understanding of gravity}
when one considers that multi-wavelength observations with space- and
ground-based instruments will allow for constraints on violations of
the strong equivalence principle that are 8 orders of magnitude more
stringent than all current bounds~\cite{Barausse:2016eii}. Single
observations with future instruments will yield constraints on the
size of a large extra-dimension (in Randall-Sundrum type models) that
are 5 orders of magnitude more stringent than current bounds with
LIGO; constraints on the temporal variability of Newton's
gravitational constant that are 12 orders of magnitude more stringent
than the best current bounds with LIGO; and constraints on the mass of
the graviton from propagation effects that are about 5 orders of
magnitude better than current bounds~\cite{Chamberlain:2017fjl},
beginning to approach the natural value of the mass of the graviton in
eV that one would expect if such a mass is connected to a solution to
the dark energy problem.

In conclusion, space-based detectors will be exceptional tools to test
the generation and propagation of GWs.  {\bf They are generally
  $2$--$4$ orders of magnitude better than current GW detectors at
  probing the generation of GWs} from binaries for theories that
produce effects at negative PN order: these include scalar-tensor
theory, Einstein-dilaton-Gauss Bonnet and dynamical Chern-Simons
gravity, theories that violate Lorentz symmetry, theories with extra
dimensions, and theories with a time-varying gravitational constant.
Space-based and third-generation Earth-based
detectors~\cite{Hall:2019xmm} can observe merging BH binaries at
cosmological distances, and their longer longer baseline will yield
{\bf tighter bounds on the propagation of GWs.}

\vspace{.1cm}

\noindent
{\bf \textit{Probing black hole spacetimes and dynamics.}}
Observational evidence for massive BHs at the centers of galaxies has
been garnered by electromagnetic radiation emitted as stars and gas
interact with the BH's gravitational potential.  Arguably the best
evidence comes from the center of our own Galaxy, where the mapping of
stars orbiting a dark, compact object of mass
$~ 4 \times 10^6\,M_\odot$ within $100$~AU has become so accurate that
it was used to detect gravitational redshift by the GRAVITY
collaboration~\cite{Abuter:2018drb}.  Space-based detectors will map
BH spacetimes down to {\bf length scales $\sim 10^4$ times smaller},
probing regions on the size of the horizon through the observation of
EMRIs.

Orbiting compact objects, such as stellar-mass BHs and neutron stars,
can probe the dark region close to the horizon of massive BHs. These
compact objects emit mHz GWs as they inspiral into the central,
massive BH.  The frequencies of the GWs emitted during the inspiral
are mostly determined by the mass of the central BH. Space-based
detectors will be sensitive to capture events from
$10^5 – 10^7 M_\odot$ BHs. EMRI signals provide opportunities to test
GR that are beyond the reach of GRAVITY or ground-based detectors. The
white paper \textit{The unique potential of extreme mass-ratio
  inspirals for gravitational-wave astronomy} by Berry et
al. describes the science enabled by these sources.  According to
current estimates, detection rates will range from a few up to a few
thousand EMRIs per year, with SNRs in the hundreds for the strongest
sources~\cite{Babak:2017tow}.

EMRI orbits exhibit complicated behavior, and this complexity can be
used -- in analogy to geodesy -- to provide exquisite measurements of
the multipolar structure of the central object's spacetime.  For
rotating (Kerr) BHs in GR, all multipoles depend on just two
parameters: the mass $M$ and dimensionless spin $\chi$ of the BH.
This means that EMRIs can be used to identify any deviations of the
spacetime from the Kerr metric predicted by general
relativity~\cite{Ryan:1995wh,Ryan:1997hg,Gair:2012nm}. For example,
every EMRI detection will provide a constraint on {\bf deviations of
  the quadrupole moment from the value predicted by the Kerr solution
  at the level of 0.01--1\%~\cite{Gair:2017ynp,Babak:2017tow}.}

Other ideas have been put forth to draw inferences on the nature of
BHs. The dynamics of hypothetical BH alternatives in binary systems
are driven by their so-called ``tidal Love numbers,'' which encode the
deformability of a self-gravitating object immersed in a tidal
environment. In GR, the tidal Love numbers of black holes are exactly
zero. Recent work computed the tidal Love numbers of exotic compact
objects (such as boson stars, gravastars and wormholes) as well as BHs
in various theories of modified gravity~\cite{Cardoso:2017cfl}.
Space-based detectors could distinguish even extremely compact exotic
objects from BHs~\cite{Maselli:2017cmm}.

The large number of orbits of a small object inspiraling into a
massive BH can reveal the nature of the central BH with great
precision.  The final phase of a BH merger also provides rich ground
for testing GR.  The BH remnant of a binary BH merger is highly
distorted, and it radiates away these distortions by vibrating like a
ringing bell in a discrete set of damped oscillation frequencies
called ``quasinormal modes.''  One can then imagine treating BHs as
``gravitational atoms,'' and viewing their GW oscillation spectrum as
a unique fingerprint of spacetime dynamics (in analogy with atomic
spectra).  This is usually called ``BH
spectroscopy''~\cite{Detweiler:1980gk}.  In general, a binary BH
merger signal will contain several ringdown modes, although one
expects the weaker modes to be hard to resolve if their amplitude is
low and/or if the detector's noise is large. In GR the mode
frequencies and damping times depend only on the Kerr BH mass and
dimensionless spin $(M,\,\chi)$. Therefore the dominant mode can be
used to identify the two numbers necessary to specify the Kerr metric;
then the detection of \textit{any} subdominant mode is a test of GR,
because all complex frequencies are uniquely determined by
$(M,\,\chi)$.  BH spectroscopy provides important tests of the degree
to which the assumptions that go into the mathematical ``no-hair''
theorems of GR are violated in astrophysical BH
mergers~\cite{Cardoso:2016ryw}, where the spacetime is highly
dynamical and non-stationary.

One of the biggest puzzles in physics is the so-called ``information
loss paradox'': if a BH evaporates away and disappears, as Hawking
predicted, it destroys the information that fell in. This violates
unitarity, a foundational principle of quantum mechanics. Among
proposed solutions to the paradox there are scenarios (including
``firewall'' and ``fuzzball'' proposals) that predict quantum
modifications at the horizon scale~\cite{Giddings:2017jts}. If a
merger remnant does not possess an event horizon, the standard
ringdown signal could be followed by quasiperiodic bursts of radiation
(``echoes'') that carry information about near-horizon structures,
which are conjectured to exist in some models of quantum
gravity~\cite{Barausse:2014tra,Cardoso:2016rao}.  Measurements of
post-merger radiation~\cite{Cardoso:2017cqb} and of stochastic GW
backgrounds~\cite{Barausse:2018vdb} with Earth- or space-based
interferometers could constrain these scenarios.

A range of GW detectors could detect the quasinormal modes of a
BH. However, third-generation ground-based detectors (like the
Einstein Telescope and Cosmic Explorer) are needed to match the SNR of
space-based detectors and to perform BH
spectroscopy~\cite{Berti:2016lat}.  An important difference between
Earth- and space-based detectors is that {\bf a very large fraction of
  BH spectroscopy tests will occur at cosmological redshift in
  space-based (but not in Earth-based) detectors}.  Even
third-generation detectors like Einstein Telescope would be limited to
$z\lesssim 3$, and only 40-km detectors, such as Cosmic
Explorer~\cite{Evans:2016mbw}, would be able to do spectroscopy at
$z\approx 10$. By contrast, BH merger SNRs in space are so large that
we could detect several modes and do BH spectroscopy out to
$z\approx 5$, $10$, or even $20$, depending only on uncertainties in
astrophysical BH formation
models~\cite{Berti:2016lat,Baibhav:2018rfk}.  This would allow
simultaneous constraints of the large-scale dynamics of gravity (which
may differ from the standard $\Lambda$CDM scenario if, say,
cosmological expansion is due to gravitational degrees of freedom that
evolve with redshift) and of the strong field, highly dynamical
regime. As a corollary of this kind of analysis, GW detectors in space
will produce an exquisitely accurate redshift survey of BH masses and
spins which is of enormous value to astronomy (beyond its intrinsic
theoretical physics interest). Quite remarkably, BH mass and spin
measurements can also be used to probe dark matter, as we discuss
below.

\vspace{.1cm}

\noindent
{\bf \textit{Probing dark matter.}}
One of the most extraordinary features of massive, rotating BHs is
that they can act as particle detectors, and therefore confirm or
rule out the existence of light bosonic fields, which have been
proposed as dark matter candidates~\cite{Marsh:2015xka,Hui:2016ltb},
in the Universe. {\bf Observations of rotating BHs with space-based GW
  detectors could therefore constrain or detect certain dark matter
  candidates, even in the absence of a direct detection of stochastic
  GWs of cosmological origin.}  The reason is that ultralight bosonic
fields around spinning BHs can trigger a superradiant instability that
Press and Teukolsky called a ``BH bomb''~\cite{Press:1972zz}, forming
a long-lived bosonic ``cloud'' outside the horizon.  The superradiant
instability spins the BH down, transferring up to a few percent of the
BH's mass and angular momentum to the
cloud~\cite{Arvanitaki:2010sy,Brito:2014wla,Arvanitaki:2014wva,Yoshino:2014wwa,Brito:2015oca,Arvanitaki:2016qwi,Baryakhtar:2017ngi,East:2017ovw}.
The condensate is then dissipated through the emission of GWs with
frequency $f\sim m_s/\hbar$, where $m_s$ is the mass of the field.
This explosive mechanism is most effective when the boson's Compton
wavelength is comparable to the BH's gravitational
radius~\cite{Dolan:2007mj}.  Strong motivation to investigate this
possibility comes e.g. from ``string axiverse'' scenarios in particle
theory (where axion-like particles arise over a broad range of masses
in string theory compactifications as Kaluza–Klein zero modes of
antisymmetric tensor fields~\cite{Arvanitaki:2009fg}) and from ``fuzzy
dark matter'' scenarios (which require axions with masses
$\approx 10^{-22}$~eV~\cite{Hui:2016ltb}).  Current Earth-based
detectors can probe boson masses $m_s\sim 10^{-13}$--$10^{-11}$~eV,
while {\bf a space-based detector can detect or rule out bosons of
  mass
  $m_s\sim
  10^{-19}$--$10^{-15}$~eV}~\cite{Brito:2017wnc,Brito:2017zvb,Isi:2018pzk,Ghosh:2018gaw,Tsukada:2018mbp}.
While axions in the ``standard'' mass range proposed to solve the
strong CP problem of QCD could be tested by GW interferometers on
Earth~\cite{Arvanitaki:2014wva,Arvanitaki:2016qwi,Brito:2017wnc}, LISA
could test a broad range of masses relevant to string axiverse
scenarios, as well as some candidates for fuzzy dark matter.

The range of allowed boson masses $m_s$ can also be constrained by
LISA measurements of the spins of BHs in binary systems.  For a given
$m_s$, spinning BHs should not exist when the BH spin $\chi$ is large
enough to trigger superradiant instabilities.  Instability windows in
the BH spin versus mass plane, for selected values of $m_s$, can be
obtained by requiring that the instability acts on timescales shorter
than known astrophysical processes, such as accretion and mergers.
Roughly speaking, continuum fitting or Iron K$\alpha$ measurements of
supermassive BH spins probe the existence of bosons in the mass range
$m_s\sim 10^{-19}$--$10^{-17}$~eV. For stellar-mass BHs, the relevant
mass range is $m_s\sim 10^{-12}$--$10^{-11}$~eV.  BH spin measurements
with a space-based GW detector can rule out light dark matter
particles in the intermediate mass range (roughly
$m_s\sim 10^{-16}$--$10^{-13}$~eV) inaccessible to electromagnetic
observations of stellar and massive BHs.  {\bf Space-based GW
  detectors can probe the existence of light scalar fields in a large
  mass range that is not probed by other BH spin measurement methods,
  or even measure $m_s$ with $\sim 10\%$ accuracy if scalars in the
  mass range $[10^{-17}, 10^{-13}]$~eV exist in
  nature}~\cite{Brito:2017zvb}.  Spin-one and spin-two fields (i.e.,
hypothetical dark photons or massive gravitons) would trigger even
stronger superradiant instabilities, so a space-based detector could
either detect them or set strong constraints on their
existence~\cite{Pani:2012vp,East:2017ovw,Brito:2013wya}.

Another interesting candidate for dark matter are primordial BHs
(PBHs) \cite{Hawking1971}. In particular, PBHs in the stellar-mass
range may contribute a non-negligible fraction of dark
matter~\cite{Bird:2016dcv,Clesse:2016vqa,Sasaki:2016jop,Sasaki:2018dmp}.
PBHs can dynamically form binaries, typically resulting in highly
eccentric orbits at formation~\cite{Ali-Haimoud:2017rtz}. GWs are a
direct probe of the self-interaction of PBH dark
matter~\cite{Kovetz:2017rvv}.  With its access to earlier stages of
the inspiral, a space-based detector can be invaluable in
distinguishing the PBH binary formation channel from stellar-origin
formation channels through measurements of the spin and
eccentricity~\cite{Cholis:2016kqi}, as well as the mass
spectrum~\cite{Kovetz:2016kpi}. Another source of unique information
is through the stochastic background. The PBH merger rate at high
redshift is not limited by the star formation rate, and so the
stochastic background from these events should extend to lower
frequencies (and higher redshifts) than for traditional binary BH
sources~\cite{Mandic:2016lcn,Clesse:2016ajp}.  Meanwhile, if PBHs are
to form from the collapse of overdense regions deep in the radiation
domination era, the required $\mathcal{O}(1)$ fluctuations in the
primordial curvature power spectrum will provide a second-order source
of primordial
GWs~\cite{Ananda:2006af,Baumann:2007zm,Garcia-Bellido:2017aan}. The
characteristic frequency of these GWs is directly related to the PBH
mass. Interestingly, {\bf one of the least constrained mass windows
  for PBH dark matter -- $10^{-13}\,M_\odot$ to $10^{-11}\,M_\odot$ --
  corresponds precisely to the mHz frequency window of
  LISA}~\cite{Bartolo:2018evs,Bartolo:2018rku,Cai:2018dig}. LISA will
be able to test the PBH dark matter scenario in this mass window
through the two-point and three-point correlations of the GW
signal~\cite{Bartolo:2016ami,Bartolo:2018qqn}.

In conclusion, space-based interferometers usher in the promise of mHz
GW astronomy and, with it, the power to test our understanding of
gravitational physics, from modifications of GR to hints at the true
nature of dark matter.  Space-based GW detectors will dramatically
advance, and potentially revolutionize, our understanding of
fundamental physics and astrophysics.

\newpage
\bibliographystyle{apsrev}

\bibliography{LISATGRWP}

\begin{thebibliography}{79}
\expandafter\ifx\csname natexlab\endcsname\relax\def\natexlab#1{#1}\fi
\expandafter\ifx\csname bibnamefont\endcsname\relax
  \def\bibnamefont#1{#1}\fi
\expandafter\ifx\csname bibfnamefont\endcsname\relax
  \def\bibfnamefont#1{#1}\fi
\expandafter\ifx\csname citenamefont\endcsname\relax
  \def\citenamefont#1{#1}\fi
\expandafter\ifx\csname url\endcsname\relax
  \def\url#1{\texttt{#1}}\fi
\expandafter\ifx\csname urlprefix\endcsname\relax\def\urlprefix{URL }\fi
\providecommand{\bibinfo}[2]{#2}
\providecommand{\eprint}[2][]{\url{#2}}

\bibitem[{\citenamefont{Hulse and Taylor}(1975)}]{Hulse:1974eb}
\bibinfo{author}{\bibfnamefont{R.~A.} \bibnamefont{Hulse}} \bibnamefont{and}
  \bibinfo{author}{\bibfnamefont{J.~H.} \bibnamefont{Taylor}},
  \bibinfo{journal}{Astrophys. J.} \textbf{\bibinfo{volume}{195}},
  \bibinfo{pages}{L51} (\bibinfo{year}{1975}).

\bibitem[{\citenamefont{Abbott et~al.}(2016{\natexlab{a}})}]{Abbott:2016blz}
\bibinfo{author}{\bibfnamefont{B.~P.} \bibnamefont{Abbott}}
  \bibnamefont{et~al.} (\bibinfo{collaboration}{Virgo, LIGO Scientific}),
  \bibinfo{journal}{Phys. Rev. Lett.} \textbf{\bibinfo{volume}{116}},
  \bibinfo{pages}{061102} (\bibinfo{year}{2016}{\natexlab{a}}),
  \eprint{1602.03837}.

\bibitem[{\citenamefont{Abramovici et~al.}(1992)}]{Abramovici:1992ah}
\bibinfo{author}{\bibfnamefont{A.}~\bibnamefont{Abramovici}}
  \bibnamefont{et~al.}, \bibinfo{journal}{Science}
  \textbf{\bibinfo{volume}{256}}, \bibinfo{pages}{325} (\bibinfo{year}{1992}).

\bibitem[{\citenamefont{Giazotto}(1990)}]{Giazotto:1988gw}
\bibinfo{author}{\bibfnamefont{A.}~\bibnamefont{Giazotto}},
  \bibinfo{journal}{Nucl. Instrum. Meth.} \textbf{\bibinfo{volume}{A289}},
  \bibinfo{pages}{518} (\bibinfo{year}{1990}).

\bibitem[{\citenamefont{Verbiest et~al.}(2016)}]{Verbiest:2016vem}
\bibinfo{author}{\bibfnamefont{J.~P.~W.} \bibnamefont{Verbiest}}
  \bibnamefont{et~al.}, \bibinfo{journal}{Mon. Not. Roy. Astron. Soc.}
  \textbf{\bibinfo{volume}{458}}, \bibinfo{pages}{1267} (\bibinfo{year}{2016}),
  \eprint{1602.03640}.

\bibitem[{\citenamefont{Audley et~al.}(2017)}]{Audley:2017drz}
\bibinfo{author}{\bibfnamefont{H.}~\bibnamefont{Audley}} \bibnamefont{et~al.}
  (\bibinfo{year}{2017}), \eprint{1702.00786}.

\bibitem[{\citenamefont{Amaro-Seoane et~al.}(2012)}]{AmaroSeoane:2012je}
\bibinfo{author}{\bibfnamefont{P.}~\bibnamefont{Amaro-Seoane}}
  \bibnamefont{et~al.}, \bibinfo{journal}{Class. Quant. Grav.}
  \textbf{\bibinfo{volume}{29}}, \bibinfo{pages}{124016}
  (\bibinfo{year}{2012}), \eprint{1202.0839}.

\bibitem[{\citenamefont{Will}(2014)}]{Will:2014kxa}
\bibinfo{author}{\bibfnamefont{C.~M.} \bibnamefont{Will}},
  \bibinfo{journal}{Living Rev. Rel.} \textbf{\bibinfo{volume}{17}},
  \bibinfo{pages}{4} (\bibinfo{year}{2014}), \eprint{1403.7377}.

\bibitem[{\citenamefont{Yunes and Siemens}(2013)}]{Yunes:2013dva}
\bibinfo{author}{\bibfnamefont{N.}~\bibnamefont{Yunes}} \bibnamefont{and}
  \bibinfo{author}{\bibfnamefont{X.}~\bibnamefont{Siemens}},
  \bibinfo{journal}{Living Rev. Rel.} \textbf{\bibinfo{volume}{16}},
  \bibinfo{pages}{9} (\bibinfo{year}{2013}), \eprint{1304.3473}.

\bibitem[{\citenamefont{Berti et~al.}(2015)}]{Berti:2015itd}
\bibinfo{author}{\bibfnamefont{E.}~\bibnamefont{Berti}} \bibnamefont{et~al.},
  \bibinfo{journal}{Class. Quant. Grav.} \textbf{\bibinfo{volume}{32}},
  \bibinfo{pages}{243001} (\bibinfo{year}{2015}), \eprint{1501.07274}.

\bibitem[{\citenamefont{Gair et~al.}(2013)\citenamefont{Gair, Vallisneri,
  Larson, and Baker}}]{Gair:2012nm}
\bibinfo{author}{\bibfnamefont{J.~R.} \bibnamefont{Gair}},
  \bibinfo{author}{\bibfnamefont{M.}~\bibnamefont{Vallisneri}},
  \bibinfo{author}{\bibfnamefont{S.~L.} \bibnamefont{Larson}},
  \bibnamefont{and} \bibinfo{author}{\bibfnamefont{J.~G.} \bibnamefont{Baker}},
  \bibinfo{journal}{Living Rev. Rel.} \textbf{\bibinfo{volume}{16}},
  \bibinfo{pages}{7} (\bibinfo{year}{2013}), \eprint{1212.5575}.

\bibitem[{\citenamefont{Berti et~al.}(2009)\citenamefont{Berti, Cardoso, and
  Starinets}}]{Berti:2009kk}
\bibinfo{author}{\bibfnamefont{E.}~\bibnamefont{Berti}},
  \bibinfo{author}{\bibfnamefont{V.}~\bibnamefont{Cardoso}}, \bibnamefont{and}
  \bibinfo{author}{\bibfnamefont{A.~O.} \bibnamefont{Starinets}},
  \bibinfo{journal}{Class. Quant. Grav.} \textbf{\bibinfo{volume}{26}},
  \bibinfo{pages}{163001} (\bibinfo{year}{2009}), \eprint{0905.2975}.

\bibitem[{\citenamefont{Nordtvedt}(1968)}]{Nordtvedt:1968qs}
\bibinfo{author}{\bibfnamefont{K.}~\bibnamefont{Nordtvedt}},
  \bibinfo{journal}{Phys. Rev.} \textbf{\bibinfo{volume}{169}},
  \bibinfo{pages}{1017} (\bibinfo{year}{1968}).

\bibitem[{\citenamefont{{Will}}(1971)}]{1971ApJ...163..611W}
\bibinfo{author}{\bibfnamefont{C.~M.} \bibnamefont{{Will}}},
  \bibinfo{journal}{\apj} \textbf{\bibinfo{volume}{163}}, \bibinfo{pages}{611}
  (\bibinfo{year}{1971}).

\bibitem[{\citenamefont{{Will} and {Nordtvedt}}(1972)}]{1972ApJ...177..757W}
\bibinfo{author}{\bibfnamefont{C.~M.} \bibnamefont{{Will}}} \bibnamefont{and}
  \bibinfo{author}{\bibfnamefont{K.}~\bibnamefont{{Nordtvedt}},
  \bibfnamefont{Jr.}}, \bibinfo{journal}{\apj} \textbf{\bibinfo{volume}{177}},
  \bibinfo{pages}{757} (\bibinfo{year}{1972}).

\bibitem[{\citenamefont{{Nordtvedt} and {Will}}(1972)}]{1972ApJ...177..775N}
\bibinfo{author}{\bibfnamefont{K.}~\bibnamefont{{Nordtvedt}},
  \bibfnamefont{Jr.}} \bibnamefont{and} \bibinfo{author}{\bibfnamefont{C.~M.}
  \bibnamefont{{Will}}}, \bibinfo{journal}{\apj}
  \textbf{\bibinfo{volume}{177}}, \bibinfo{pages}{775} (\bibinfo{year}{1972}).

\bibitem[{\citenamefont{Abbott
  et~al.}(2016{\natexlab{b}})}]{TheLIGOScientific:2016src}
\bibinfo{author}{\bibfnamefont{B.~P.} \bibnamefont{Abbott}}
  \bibnamefont{et~al.} (\bibinfo{collaboration}{Virgo, LIGO Scientific}),
  \bibinfo{journal}{Phys. Rev. Lett.} \textbf{\bibinfo{volume}{116}},
  \bibinfo{pages}{221101} (\bibinfo{year}{2016}{\natexlab{b}}),
  \eprint{1602.03841}.

\bibitem[{\citenamefont{Arun et~al.}(2006)\citenamefont{Arun, Iyer, Qusailah,
  and Sathyaprakash}}]{Arun:2006yw}
\bibinfo{author}{\bibfnamefont{K.~G.} \bibnamefont{Arun}},
  \bibinfo{author}{\bibfnamefont{B.~R.} \bibnamefont{Iyer}},
  \bibinfo{author}{\bibfnamefont{M.~S.~S.} \bibnamefont{Qusailah}},
  \bibnamefont{and} \bibinfo{author}{\bibfnamefont{B.~S.}
  \bibnamefont{Sathyaprakash}}, \bibinfo{journal}{Class. Quant. Grav.}
  \textbf{\bibinfo{volume}{23}}, \bibinfo{pages}{L37} (\bibinfo{year}{2006}),
  \eprint{gr-qc/0604018}.

\bibitem[{\citenamefont{Yunes and Pretorius}(2009)}]{Yunes:2009ke}
\bibinfo{author}{\bibfnamefont{N.}~\bibnamefont{Yunes}} \bibnamefont{and}
  \bibinfo{author}{\bibfnamefont{F.}~\bibnamefont{Pretorius}},
  \bibinfo{journal}{Phys. Rev.} \textbf{\bibinfo{volume}{D80}},
  \bibinfo{pages}{122003} (\bibinfo{year}{2009}), \eprint{0909.3328}.

\bibitem[{\citenamefont{Berti et~al.}(2018)\citenamefont{Berti, Yagi, and
  Yunes}}]{Berti:2018cxi}
\bibinfo{author}{\bibfnamefont{E.}~\bibnamefont{Berti}},
  \bibinfo{author}{\bibfnamefont{K.}~\bibnamefont{Yagi}}, \bibnamefont{and}
  \bibinfo{author}{\bibfnamefont{N.}~\bibnamefont{Yunes}}
  (\bibinfo{year}{2018}), \eprint{1801.03208}.

\bibitem[{\citenamefont{Barausse et~al.}(2016)\citenamefont{Barausse, Yunes,
  and Chamberlain}}]{Barausse:2016eii}
\bibinfo{author}{\bibfnamefont{E.}~\bibnamefont{Barausse}},
  \bibinfo{author}{\bibfnamefont{N.}~\bibnamefont{Yunes}}, \bibnamefont{and}
  \bibinfo{author}{\bibfnamefont{K.}~\bibnamefont{Chamberlain}},
  \bibinfo{journal}{Phys. Rev. Lett.} \textbf{\bibinfo{volume}{116}},
  \bibinfo{pages}{241104} (\bibinfo{year}{2016}), \eprint{1603.04075}.

\bibitem[{\citenamefont{Chamberlain and Yunes}(2017)}]{Chamberlain:2017fjl}
\bibinfo{author}{\bibfnamefont{K.}~\bibnamefont{Chamberlain}} \bibnamefont{and}
  \bibinfo{author}{\bibfnamefont{N.}~\bibnamefont{Yunes}},
  \bibinfo{journal}{Phys. Rev.} \textbf{\bibinfo{volume}{D96}},
  \bibinfo{pages}{084039} (\bibinfo{year}{2017}), \eprint{1704.08268}.

\bibitem[{\citenamefont{Hall and Evans}(2019)}]{Hall:2019xmm}
\bibinfo{author}{\bibfnamefont{E.~D.} \bibnamefont{Hall}} \bibnamefont{and}
  \bibinfo{author}{\bibfnamefont{M.}~\bibnamefont{Evans}}
  (\bibinfo{year}{2019}), \eprint{1902.09485}.

\bibitem[{\citenamefont{Abuter et~al.}(2018)}]{Abuter:2018drb}
\bibinfo{author}{\bibfnamefont{R.}~\bibnamefont{Abuter}} \bibnamefont{et~al.}
  (\bibinfo{collaboration}{GRAVITY}) (\bibinfo{year}{2018}),
  \eprint{1807.09409}.

\bibitem[{\citenamefont{Babak et~al.}(2017)\citenamefont{Babak, Gair, Sesana,
  Barausse, Sopuerta, Berry, Berti, Amaro-Seoane, Petiteau, and
  Klein}}]{Babak:2017tow}
\bibinfo{author}{\bibfnamefont{S.}~\bibnamefont{Babak}},
  \bibinfo{author}{\bibfnamefont{J.}~\bibnamefont{Gair}},
  \bibinfo{author}{\bibfnamefont{A.}~\bibnamefont{Sesana}},
  \bibinfo{author}{\bibfnamefont{E.}~\bibnamefont{Barausse}},
  \bibinfo{author}{\bibfnamefont{C.~F.} \bibnamefont{Sopuerta}},
  \bibinfo{author}{\bibfnamefont{C.~P.~L.} \bibnamefont{Berry}},
  \bibinfo{author}{\bibfnamefont{E.}~\bibnamefont{Berti}},
  \bibinfo{author}{\bibfnamefont{P.}~\bibnamefont{Amaro-Seoane}},
  \bibinfo{author}{\bibfnamefont{A.}~\bibnamefont{Petiteau}}, \bibnamefont{and}
  \bibinfo{author}{\bibfnamefont{A.}~\bibnamefont{Klein}},
  \bibinfo{journal}{Phys. Rev.} \textbf{\bibinfo{volume}{D95}},
  \bibinfo{pages}{103012} (\bibinfo{year}{2017}), \eprint{1703.09722}.

\bibitem[{\citenamefont{Ryan}(1995)}]{Ryan:1995wh}
\bibinfo{author}{\bibfnamefont{F.~D.} \bibnamefont{Ryan}},
  \bibinfo{journal}{Phys. Rev.} \textbf{\bibinfo{volume}{D52}},
  \bibinfo{pages}{5707} (\bibinfo{year}{1995}).

\bibitem[{\citenamefont{Ryan}(1997)}]{Ryan:1997hg}
\bibinfo{author}{\bibfnamefont{F.~D.} \bibnamefont{Ryan}},
  \bibinfo{journal}{Phys. Rev.} \textbf{\bibinfo{volume}{D56}},
  \bibinfo{pages}{1845} (\bibinfo{year}{1997}).

\bibitem[{\citenamefont{Gair et~al.}(2017)\citenamefont{Gair, Babak, Sesana,
  Amaro-Seoane, Barausse, Berry, Berti, and Sopuerta}}]{Gair:2017ynp}
\bibinfo{author}{\bibfnamefont{J.~R.} \bibnamefont{Gair}},
  \bibinfo{author}{\bibfnamefont{S.}~\bibnamefont{Babak}},
  \bibinfo{author}{\bibfnamefont{A.}~\bibnamefont{Sesana}},
  \bibinfo{author}{\bibfnamefont{P.}~\bibnamefont{Amaro-Seoane}},
  \bibinfo{author}{\bibfnamefont{E.}~\bibnamefont{Barausse}},
  \bibinfo{author}{\bibfnamefont{C.~P.~L.} \bibnamefont{Berry}},
  \bibinfo{author}{\bibfnamefont{E.}~\bibnamefont{Berti}}, \bibnamefont{and}
  \bibinfo{author}{\bibfnamefont{C.}~\bibnamefont{Sopuerta}},
  \bibinfo{journal}{J. Phys. Conf. Ser.} \textbf{\bibinfo{volume}{840}},
  \bibinfo{pages}{012021} (\bibinfo{year}{2017}), \eprint{1704.00009}.

\bibitem[{\citenamefont{Cardoso et~al.}(2017)\citenamefont{Cardoso, Franzin,
  Maselli, Pani, and Raposo}}]{Cardoso:2017cfl}
\bibinfo{author}{\bibfnamefont{V.}~\bibnamefont{Cardoso}},
  \bibinfo{author}{\bibfnamefont{E.}~\bibnamefont{Franzin}},
  \bibinfo{author}{\bibfnamefont{A.}~\bibnamefont{Maselli}},
  \bibinfo{author}{\bibfnamefont{P.}~\bibnamefont{Pani}}, \bibnamefont{and}
  \bibinfo{author}{\bibfnamefont{G.}~\bibnamefont{Raposo}},
  \bibinfo{journal}{Phys. Rev.} \textbf{\bibinfo{volume}{D95}},
  \bibinfo{pages}{084014} (\bibinfo{year}{2017}), \bibinfo{note}{[Addendum:
  Phys. Rev.D95,no.8,089901(2017)]}, \eprint{1701.01116}.

\bibitem[{\citenamefont{Maselli et~al.}(2018)\citenamefont{Maselli, Pani,
  Cardoso, Abdelsalhin, Gualtieri, and Ferrari}}]{Maselli:2017cmm}
\bibinfo{author}{\bibfnamefont{A.}~\bibnamefont{Maselli}},
  \bibinfo{author}{\bibfnamefont{P.}~\bibnamefont{Pani}},
  \bibinfo{author}{\bibfnamefont{V.}~\bibnamefont{Cardoso}},
  \bibinfo{author}{\bibfnamefont{T.}~\bibnamefont{Abdelsalhin}},
  \bibinfo{author}{\bibfnamefont{L.}~\bibnamefont{Gualtieri}},
  \bibnamefont{and} \bibinfo{author}{\bibfnamefont{V.}~\bibnamefont{Ferrari}},
  \bibinfo{journal}{Phys. Rev. Lett.} \textbf{\bibinfo{volume}{120}},
  \bibinfo{pages}{081101} (\bibinfo{year}{2018}), \eprint{1703.10612}.

\bibitem[{\citenamefont{Detweiler}(1980)}]{Detweiler:1980gk}
\bibinfo{author}{\bibfnamefont{S.~L.} \bibnamefont{Detweiler}},
  \bibinfo{journal}{Astrophys. J.} \textbf{\bibinfo{volume}{239}},
  \bibinfo{pages}{292} (\bibinfo{year}{1980}).

\bibitem[{\citenamefont{Cardoso and Gualtieri}(2016)}]{Cardoso:2016ryw}
\bibinfo{author}{\bibfnamefont{V.}~\bibnamefont{Cardoso}} \bibnamefont{and}
  \bibinfo{author}{\bibfnamefont{L.}~\bibnamefont{Gualtieri}},
  \bibinfo{journal}{Class. Quant. Grav.} \textbf{\bibinfo{volume}{33}},
  \bibinfo{pages}{174001} (\bibinfo{year}{2016}), \eprint{1607.03133}.

\bibitem[{\citenamefont{Giddings}(2017)}]{Giddings:2017jts}
\bibinfo{author}{\bibfnamefont{S.~B.} \bibnamefont{Giddings}}
  (\bibinfo{year}{2017}), \eprint{1703.03387}.

\bibitem[{\citenamefont{Barausse et~al.}(2014)\citenamefont{Barausse, Cardoso,
  and Pani}}]{Barausse:2014tra}
\bibinfo{author}{\bibfnamefont{E.}~\bibnamefont{Barausse}},
  \bibinfo{author}{\bibfnamefont{V.}~\bibnamefont{Cardoso}}, \bibnamefont{and}
  \bibinfo{author}{\bibfnamefont{P.}~\bibnamefont{Pani}},
  \bibinfo{journal}{Phys. Rev.} \textbf{\bibinfo{volume}{D89}},
  \bibinfo{pages}{104059} (\bibinfo{year}{2014}), \eprint{1404.7149}.

\bibitem[{\citenamefont{Cardoso et~al.}(2016)\citenamefont{Cardoso, Franzin,
  and Pani}}]{Cardoso:2016rao}
\bibinfo{author}{\bibfnamefont{V.}~\bibnamefont{Cardoso}},
  \bibinfo{author}{\bibfnamefont{E.}~\bibnamefont{Franzin}}, \bibnamefont{and}
  \bibinfo{author}{\bibfnamefont{P.}~\bibnamefont{Pani}},
  \bibinfo{journal}{Phys. Rev. Lett.} \textbf{\bibinfo{volume}{116}},
  \bibinfo{pages}{171101} (\bibinfo{year}{2016}), \bibinfo{note}{[Erratum:
  Phys. Rev. Lett.117,no.8,089902(2016)]}, \eprint{1602.07309}.

\bibitem[{\citenamefont{Cardoso and Pani}(2017)}]{Cardoso:2017cqb}
\bibinfo{author}{\bibfnamefont{V.}~\bibnamefont{Cardoso}} \bibnamefont{and}
  \bibinfo{author}{\bibfnamefont{P.}~\bibnamefont{Pani}},
  \bibinfo{journal}{Nat. Astron.} \textbf{\bibinfo{volume}{1}},
  \bibinfo{pages}{586} (\bibinfo{year}{2017}), \eprint{1709.01525}.

\bibitem[{\citenamefont{Barausse et~al.}(2018)\citenamefont{Barausse, Brito,
  Cardoso, Dvorkin, and Pani}}]{Barausse:2018vdb}
\bibinfo{author}{\bibfnamefont{E.}~\bibnamefont{Barausse}},
  \bibinfo{author}{\bibfnamefont{R.}~\bibnamefont{Brito}},
  \bibinfo{author}{\bibfnamefont{V.}~\bibnamefont{Cardoso}},
  \bibinfo{author}{\bibfnamefont{I.}~\bibnamefont{Dvorkin}}, \bibnamefont{and}
  \bibinfo{author}{\bibfnamefont{P.}~\bibnamefont{Pani}},
  \bibinfo{journal}{Class. Quant. Grav.} \textbf{\bibinfo{volume}{35}},
  \bibinfo{pages}{20LT01} (\bibinfo{year}{2018}), \eprint{1805.08229}.

\bibitem[{\citenamefont{Berti et~al.}(2016)\citenamefont{Berti, Sesana,
  Barausse, Cardoso, and Belczynski}}]{Berti:2016lat}
\bibinfo{author}{\bibfnamefont{E.}~\bibnamefont{Berti}},
  \bibinfo{author}{\bibfnamefont{A.}~\bibnamefont{Sesana}},
  \bibinfo{author}{\bibfnamefont{E.}~\bibnamefont{Barausse}},
  \bibinfo{author}{\bibfnamefont{V.}~\bibnamefont{Cardoso}}, \bibnamefont{and}
  \bibinfo{author}{\bibfnamefont{K.}~\bibnamefont{Belczynski}},
  \bibinfo{journal}{Phys. Rev. Lett.} \textbf{\bibinfo{volume}{117}},
  \bibinfo{pages}{101102} (\bibinfo{year}{2016}), \eprint{1605.09286}.

\bibitem[{\citenamefont{Abbott et~al.}(2017)}]{Evans:2016mbw}
\bibinfo{author}{\bibfnamefont{B.~P.} \bibnamefont{Abbott}}
  \bibnamefont{et~al.} (\bibinfo{collaboration}{LIGO Scientific}),
  \bibinfo{journal}{Class. Quant. Grav.} \textbf{\bibinfo{volume}{34}},
  \bibinfo{pages}{044001} (\bibinfo{year}{2017}), \eprint{1607.08697}.

\bibitem[{\citenamefont{Baibhav and Berti}(2018)}]{Baibhav:2018rfk}
\bibinfo{author}{\bibfnamefont{V.}~\bibnamefont{Baibhav}} \bibnamefont{and}
  \bibinfo{author}{\bibfnamefont{E.}~\bibnamefont{Berti}}
  (\bibinfo{year}{2018}), \eprint{1809.03500}.

\bibitem[{\citenamefont{Marsh}(2016)}]{Marsh:2015xka}
\bibinfo{author}{\bibfnamefont{D.~J.~E.} \bibnamefont{Marsh}},
  \bibinfo{journal}{Phys. Rept.} \textbf{\bibinfo{volume}{643}},
  \bibinfo{pages}{1} (\bibinfo{year}{2016}), \eprint{1510.07633}.

\bibitem[{\citenamefont{Hui et~al.}(2017)\citenamefont{Hui, Ostriker, Tremaine,
  and Witten}}]{Hui:2016ltb}
\bibinfo{author}{\bibfnamefont{L.}~\bibnamefont{Hui}},
  \bibinfo{author}{\bibfnamefont{J.~P.} \bibnamefont{Ostriker}},
  \bibinfo{author}{\bibfnamefont{S.}~\bibnamefont{Tremaine}}, \bibnamefont{and}
  \bibinfo{author}{\bibfnamefont{E.}~\bibnamefont{Witten}},
  \bibinfo{journal}{Phys. Rev.} \textbf{\bibinfo{volume}{D95}},
  \bibinfo{pages}{043541} (\bibinfo{year}{2017}), \eprint{1610.08297}.

\bibitem[{\citenamefont{Press and Teukolsky}(1972)}]{Press:1972zz}
\bibinfo{author}{\bibfnamefont{W.~H.} \bibnamefont{Press}} \bibnamefont{and}
  \bibinfo{author}{\bibfnamefont{S.~A.} \bibnamefont{Teukolsky}},
  \bibinfo{journal}{Nature} \textbf{\bibinfo{volume}{238}},
  \bibinfo{pages}{211} (\bibinfo{year}{1972}).

\bibitem[{\citenamefont{Arvanitaki and Dubovsky}(2011)}]{Arvanitaki:2010sy}
\bibinfo{author}{\bibfnamefont{A.}~\bibnamefont{Arvanitaki}} \bibnamefont{and}
  \bibinfo{author}{\bibfnamefont{S.}~\bibnamefont{Dubovsky}},
  \bibinfo{journal}{Phys. Rev.} \textbf{\bibinfo{volume}{D83}},
  \bibinfo{pages}{044026} (\bibinfo{year}{2011}), \eprint{1004.3558}.

\bibitem[{\citenamefont{Brito et~al.}(2015{\natexlab{a}})\citenamefont{Brito,
  Cardoso, and Pani}}]{Brito:2014wla}
\bibinfo{author}{\bibfnamefont{R.}~\bibnamefont{Brito}},
  \bibinfo{author}{\bibfnamefont{V.}~\bibnamefont{Cardoso}}, \bibnamefont{and}
  \bibinfo{author}{\bibfnamefont{P.}~\bibnamefont{Pani}},
  \bibinfo{journal}{Class. Quant. Grav.} \textbf{\bibinfo{volume}{32}},
  \bibinfo{pages}{134001} (\bibinfo{year}{2015}{\natexlab{a}}),
  \eprint{1411.0686}.

\bibitem[{\citenamefont{Arvanitaki et~al.}(2015)\citenamefont{Arvanitaki,
  Baryakhtar, and Huang}}]{Arvanitaki:2014wva}
\bibinfo{author}{\bibfnamefont{A.}~\bibnamefont{Arvanitaki}},
  \bibinfo{author}{\bibfnamefont{M.}~\bibnamefont{Baryakhtar}},
  \bibnamefont{and} \bibinfo{author}{\bibfnamefont{X.}~\bibnamefont{Huang}},
  \bibinfo{journal}{Phys. Rev.} \textbf{\bibinfo{volume}{D91}},
  \bibinfo{pages}{084011} (\bibinfo{year}{2015}), \eprint{1411.2263}.

\bibitem[{\citenamefont{Yoshino and Kodama}(2015)}]{Yoshino:2014wwa}
\bibinfo{author}{\bibfnamefont{H.}~\bibnamefont{Yoshino}} \bibnamefont{and}
  \bibinfo{author}{\bibfnamefont{H.}~\bibnamefont{Kodama}},
  \bibinfo{journal}{PTEP} \textbf{\bibinfo{volume}{2015}},
  \bibinfo{pages}{061E01} (\bibinfo{year}{2015}), \eprint{1407.2030}.

\bibitem[{\citenamefont{Brito et~al.}(2015{\natexlab{b}})\citenamefont{Brito,
  Cardoso, and Pani}}]{Brito:2015oca}
\bibinfo{author}{\bibfnamefont{R.}~\bibnamefont{Brito}},
  \bibinfo{author}{\bibfnamefont{V.}~\bibnamefont{Cardoso}}, \bibnamefont{and}
  \bibinfo{author}{\bibfnamefont{P.}~\bibnamefont{Pani}},
  \bibinfo{journal}{Lect. Notes Phys.} \textbf{\bibinfo{volume}{906}},
  \bibinfo{pages}{pp.1} (\bibinfo{year}{2015}{\natexlab{b}}),
  \eprint{1501.06570}.

\bibitem[{\citenamefont{Arvanitaki et~al.}(2017)\citenamefont{Arvanitaki,
  Baryakhtar, Dimopoulos, Dubovsky, and Lasenby}}]{Arvanitaki:2016qwi}
\bibinfo{author}{\bibfnamefont{A.}~\bibnamefont{Arvanitaki}},
  \bibinfo{author}{\bibfnamefont{M.}~\bibnamefont{Baryakhtar}},
  \bibinfo{author}{\bibfnamefont{S.}~\bibnamefont{Dimopoulos}},
  \bibinfo{author}{\bibfnamefont{S.}~\bibnamefont{Dubovsky}}, \bibnamefont{and}
  \bibinfo{author}{\bibfnamefont{R.}~\bibnamefont{Lasenby}},
  \bibinfo{journal}{Phys. Rev.} \textbf{\bibinfo{volume}{D95}},
  \bibinfo{pages}{043001} (\bibinfo{year}{2017}), \eprint{1604.03958}.

\bibitem[{\citenamefont{Baryakhtar et~al.}(2017)\citenamefont{Baryakhtar,
  Lasenby, and Teo}}]{Baryakhtar:2017ngi}
\bibinfo{author}{\bibfnamefont{M.}~\bibnamefont{Baryakhtar}},
  \bibinfo{author}{\bibfnamefont{R.}~\bibnamefont{Lasenby}}, \bibnamefont{and}
  \bibinfo{author}{\bibfnamefont{M.}~\bibnamefont{Teo}},
  \bibinfo{journal}{Phys. Rev.} \textbf{\bibinfo{volume}{D96}},
  \bibinfo{pages}{035019} (\bibinfo{year}{2017}), \eprint{1704.05081}.

\bibitem[{\citenamefont{East and Pretorius}(2017)}]{East:2017ovw}
\bibinfo{author}{\bibfnamefont{W.~E.} \bibnamefont{East}} \bibnamefont{and}
  \bibinfo{author}{\bibfnamefont{F.}~\bibnamefont{Pretorius}},
  \bibinfo{journal}{Phys. Rev. Lett.} \textbf{\bibinfo{volume}{119}},
  \bibinfo{pages}{041101} (\bibinfo{year}{2017}), \eprint{1704.04791}.

\bibitem[{\citenamefont{Dolan}(2007)}]{Dolan:2007mj}
\bibinfo{author}{\bibfnamefont{S.~R.} \bibnamefont{Dolan}},
  \bibinfo{journal}{Phys. Rev.} \textbf{\bibinfo{volume}{D76}},
  \bibinfo{pages}{084001} (\bibinfo{year}{2007}), \eprint{0705.2880}.

\bibitem[{\citenamefont{Arvanitaki et~al.}(2010)\citenamefont{Arvanitaki,
  Dimopoulos, Dubovsky, Kaloper, and March-Russell}}]{Arvanitaki:2009fg}
\bibinfo{author}{\bibfnamefont{A.}~\bibnamefont{Arvanitaki}},
  \bibinfo{author}{\bibfnamefont{S.}~\bibnamefont{Dimopoulos}},
  \bibinfo{author}{\bibfnamefont{S.}~\bibnamefont{Dubovsky}},
  \bibinfo{author}{\bibfnamefont{N.}~\bibnamefont{Kaloper}}, \bibnamefont{and}
  \bibinfo{author}{\bibfnamefont{J.}~\bibnamefont{March-Russell}},
  \bibinfo{journal}{Phys. Rev.} \textbf{\bibinfo{volume}{D81}},
  \bibinfo{pages}{123530} (\bibinfo{year}{2010}), \eprint{0905.4720}.

\bibitem[{\citenamefont{Brito et~al.}(2017{\natexlab{a}})\citenamefont{Brito,
  Ghosh, Barausse, Berti, Cardoso, Dvorkin, Klein, and Pani}}]{Brito:2017wnc}
\bibinfo{author}{\bibfnamefont{R.}~\bibnamefont{Brito}},
  \bibinfo{author}{\bibfnamefont{S.}~\bibnamefont{Ghosh}},
  \bibinfo{author}{\bibfnamefont{E.}~\bibnamefont{Barausse}},
  \bibinfo{author}{\bibfnamefont{E.}~\bibnamefont{Berti}},
  \bibinfo{author}{\bibfnamefont{V.}~\bibnamefont{Cardoso}},
  \bibinfo{author}{\bibfnamefont{I.}~\bibnamefont{Dvorkin}},
  \bibinfo{author}{\bibfnamefont{A.}~\bibnamefont{Klein}}, \bibnamefont{and}
  \bibinfo{author}{\bibfnamefont{P.}~\bibnamefont{Pani}},
  \bibinfo{journal}{Phys. Rev. Lett.} \textbf{\bibinfo{volume}{119}},
  \bibinfo{pages}{131101} (\bibinfo{year}{2017}{\natexlab{a}}),
  \eprint{1706.05097}.

\bibitem[{\citenamefont{Brito et~al.}(2017{\natexlab{b}})\citenamefont{Brito,
  Ghosh, Barausse, Berti, Cardoso, Dvorkin, Klein, and Pani}}]{Brito:2017zvb}
\bibinfo{author}{\bibfnamefont{R.}~\bibnamefont{Brito}},
  \bibinfo{author}{\bibfnamefont{S.}~\bibnamefont{Ghosh}},
  \bibinfo{author}{\bibfnamefont{E.}~\bibnamefont{Barausse}},
  \bibinfo{author}{\bibfnamefont{E.}~\bibnamefont{Berti}},
  \bibinfo{author}{\bibfnamefont{V.}~\bibnamefont{Cardoso}},
  \bibinfo{author}{\bibfnamefont{I.}~\bibnamefont{Dvorkin}},
  \bibinfo{author}{\bibfnamefont{A.}~\bibnamefont{Klein}}, \bibnamefont{and}
  \bibinfo{author}{\bibfnamefont{P.}~\bibnamefont{Pani}},
  \bibinfo{journal}{Phys. Rev.} \textbf{\bibinfo{volume}{D96}},
  \bibinfo{pages}{064050} (\bibinfo{year}{2017}{\natexlab{b}}),
  \eprint{1706.06311}.

\bibitem[{\citenamefont{Isi et~al.}(2018)\citenamefont{Isi, Sun, Brito, and
  Melatos}}]{Isi:2018pzk}
\bibinfo{author}{\bibfnamefont{M.}~\bibnamefont{Isi}},
  \bibinfo{author}{\bibfnamefont{L.}~\bibnamefont{Sun}},
  \bibinfo{author}{\bibfnamefont{R.}~\bibnamefont{Brito}}, \bibnamefont{and}
  \bibinfo{author}{\bibfnamefont{A.}~\bibnamefont{Melatos}}
  (\bibinfo{year}{2018}), \eprint{1810.03812}.

\bibitem[{\citenamefont{Ghosh et~al.}(2018)\citenamefont{Ghosh, Berti, Brito,
  and Richartz}}]{Ghosh:2018gaw}
\bibinfo{author}{\bibfnamefont{S.}~\bibnamefont{Ghosh}},
  \bibinfo{author}{\bibfnamefont{E.}~\bibnamefont{Berti}},
  \bibinfo{author}{\bibfnamefont{R.}~\bibnamefont{Brito}}, \bibnamefont{and}
  \bibinfo{author}{\bibfnamefont{M.}~\bibnamefont{Richartz}}
  (\bibinfo{year}{2018}), \eprint{1812.01620}.

\bibitem[{\citenamefont{Tsukada et~al.}(2018)\citenamefont{Tsukada, Callister,
  Matas, and Meyers}}]{Tsukada:2018mbp}
\bibinfo{author}{\bibfnamefont{L.}~\bibnamefont{Tsukada}},
  \bibinfo{author}{\bibfnamefont{T.}~\bibnamefont{Callister}},
  \bibinfo{author}{\bibfnamefont{A.}~\bibnamefont{Matas}}, \bibnamefont{and}
  \bibinfo{author}{\bibfnamefont{P.}~\bibnamefont{Meyers}}
  (\bibinfo{year}{2018}), \eprint{1812.09622}.

\bibitem[{\citenamefont{Pani et~al.}(2012)\citenamefont{Pani, Cardoso,
  Gualtieri, Berti, and Ishibashi}}]{Pani:2012vp}
\bibinfo{author}{\bibfnamefont{P.}~\bibnamefont{Pani}},
  \bibinfo{author}{\bibfnamefont{V.}~\bibnamefont{Cardoso}},
  \bibinfo{author}{\bibfnamefont{L.}~\bibnamefont{Gualtieri}},
  \bibinfo{author}{\bibfnamefont{E.}~\bibnamefont{Berti}}, \bibnamefont{and}
  \bibinfo{author}{\bibfnamefont{A.}~\bibnamefont{Ishibashi}},
  \bibinfo{journal}{Phys. Rev. Lett.} \textbf{\bibinfo{volume}{109}},
  \bibinfo{pages}{131102} (\bibinfo{year}{2012}), \eprint{1209.0465}.

\bibitem[{\citenamefont{Brito et~al.}(2013)\citenamefont{Brito, Cardoso, and
  Pani}}]{Brito:2013wya}
\bibinfo{author}{\bibfnamefont{R.}~\bibnamefont{Brito}},
  \bibinfo{author}{\bibfnamefont{V.}~\bibnamefont{Cardoso}}, \bibnamefont{and}
  \bibinfo{author}{\bibfnamefont{P.}~\bibnamefont{Pani}},
  \bibinfo{journal}{Phys. Rev.} \textbf{\bibinfo{volume}{D88}},
  \bibinfo{pages}{023514} (\bibinfo{year}{2013}), \eprint{1304.6725}.

\bibitem[{\citenamefont{{Hawking}}(1971)}]{Hawking1971}
\bibinfo{author}{\bibfnamefont{S.}~\bibnamefont{{Hawking}}},
  \bibinfo{journal}{\mnras} \textbf{\bibinfo{volume}{152}}, \bibinfo{pages}{75}
  (\bibinfo{year}{1971}).

\bibitem[{\citenamefont{Bird et~al.}(2016)\citenamefont{Bird, Cholis, Muñoz,
  Ali-Haïmoud, Kamionkowski, Kovetz, Raccanelli, and Riess}}]{Bird:2016dcv}
\bibinfo{author}{\bibfnamefont{S.}~\bibnamefont{Bird}},
  \bibinfo{author}{\bibfnamefont{I.}~\bibnamefont{Cholis}},
  \bibinfo{author}{\bibfnamefont{J.~B.} \bibnamefont{Muñoz}},
  \bibinfo{author}{\bibfnamefont{Y.}~\bibnamefont{Ali-Haïmoud}},
  \bibinfo{author}{\bibfnamefont{M.}~\bibnamefont{Kamionkowski}},
  \bibinfo{author}{\bibfnamefont{E.~D.} \bibnamefont{Kovetz}},
  \bibinfo{author}{\bibfnamefont{A.}~\bibnamefont{Raccanelli}},
  \bibnamefont{and} \bibinfo{author}{\bibfnamefont{A.~G.} \bibnamefont{Riess}},
  \bibinfo{journal}{Phys. Rev. Lett.} \textbf{\bibinfo{volume}{116}},
  \bibinfo{pages}{201301} (\bibinfo{year}{2016}), \eprint{1603.00464}.

\bibitem[{\citenamefont{Clesse and
  García-Bellido}(2017{\natexlab{a}})}]{Clesse:2016vqa}
\bibinfo{author}{\bibfnamefont{S.}~\bibnamefont{Clesse}} \bibnamefont{and}
  \bibinfo{author}{\bibfnamefont{J.}~\bibnamefont{García-Bellido}},
  \bibinfo{journal}{Phys. Dark Univ.} \textbf{\bibinfo{volume}{15}},
  \bibinfo{pages}{142} (\bibinfo{year}{2017}{\natexlab{a}}),
  \eprint{1603.05234}.

\bibitem[{\citenamefont{Sasaki et~al.}(2016)\citenamefont{Sasaki, Suyama,
  Tanaka, and Yokoyama}}]{Sasaki:2016jop}
\bibinfo{author}{\bibfnamefont{M.}~\bibnamefont{Sasaki}},
  \bibinfo{author}{\bibfnamefont{T.}~\bibnamefont{Suyama}},
  \bibinfo{author}{\bibfnamefont{T.}~\bibnamefont{Tanaka}}, \bibnamefont{and}
  \bibinfo{author}{\bibfnamefont{S.}~\bibnamefont{Yokoyama}},
  \bibinfo{journal}{Phys. Rev. Lett.} \textbf{\bibinfo{volume}{117}},
  \bibinfo{pages}{061101} (\bibinfo{year}{2016}), \bibinfo{note}{[Erratum:
  Phys. Rev. Lett.121,no.5,059901(2018)]}, \eprint{1603.08338}.

\bibitem[{\citenamefont{Sasaki et~al.}(2018)\citenamefont{Sasaki, Suyama,
  Tanaka, and Yokoyama}}]{Sasaki:2018dmp}
\bibinfo{author}{\bibfnamefont{M.}~\bibnamefont{Sasaki}},
  \bibinfo{author}{\bibfnamefont{T.}~\bibnamefont{Suyama}},
  \bibinfo{author}{\bibfnamefont{T.}~\bibnamefont{Tanaka}}, \bibnamefont{and}
  \bibinfo{author}{\bibfnamefont{S.}~\bibnamefont{Yokoyama}},
  \bibinfo{journal}{Class. Quant. Grav.} \textbf{\bibinfo{volume}{35}},
  \bibinfo{pages}{063001} (\bibinfo{year}{2018}), \eprint{1801.05235}.

\bibitem[{\citenamefont{Ali-Haïmoud et~al.}(2017)\citenamefont{Ali-Haïmoud,
  Kovetz, and Kamionkowski}}]{Ali-Haimoud:2017rtz}
\bibinfo{author}{\bibfnamefont{Y.}~\bibnamefont{Ali-Haïmoud}},
  \bibinfo{author}{\bibfnamefont{E.~D.} \bibnamefont{Kovetz}},
  \bibnamefont{and}
  \bibinfo{author}{\bibfnamefont{M.}~\bibnamefont{Kamionkowski}},
  \bibinfo{journal}{Phys. Rev.} \textbf{\bibinfo{volume}{D96}},
  \bibinfo{pages}{123523} (\bibinfo{year}{2017}), \eprint{1709.06576}.

\bibitem[{\citenamefont{Kovetz}(2017)}]{Kovetz:2017rvv}
\bibinfo{author}{\bibfnamefont{E.~D.} \bibnamefont{Kovetz}},
  \bibinfo{journal}{Phys. Rev. Lett.} \textbf{\bibinfo{volume}{119}},
  \bibinfo{pages}{131301} (\bibinfo{year}{2017}), \eprint{1705.09182}.

\bibitem[{\citenamefont{Cholis et~al.}(2016)\citenamefont{Cholis, Kovetz,
  Ali-Haïmoud, Bird, Kamionkowski, Muñoz, and Raccanelli}}]{Cholis:2016kqi}
\bibinfo{author}{\bibfnamefont{I.}~\bibnamefont{Cholis}},
  \bibinfo{author}{\bibfnamefont{E.~D.} \bibnamefont{Kovetz}},
  \bibinfo{author}{\bibfnamefont{Y.}~\bibnamefont{Ali-Haïmoud}},
  \bibinfo{author}{\bibfnamefont{S.}~\bibnamefont{Bird}},
  \bibinfo{author}{\bibfnamefont{M.}~\bibnamefont{Kamionkowski}},
  \bibinfo{author}{\bibfnamefont{J.~B.} \bibnamefont{Muñoz}},
  \bibnamefont{and}
  \bibinfo{author}{\bibfnamefont{A.}~\bibnamefont{Raccanelli}},
  \bibinfo{journal}{Phys. Rev.} \textbf{\bibinfo{volume}{D94}},
  \bibinfo{pages}{084013} (\bibinfo{year}{2016}), \eprint{1606.07437}.

\bibitem[{\citenamefont{Kovetz et~al.}(2017)\citenamefont{Kovetz, Cholis,
  Breysse, and Kamionkowski}}]{Kovetz:2016kpi}
\bibinfo{author}{\bibfnamefont{E.~D.} \bibnamefont{Kovetz}},
  \bibinfo{author}{\bibfnamefont{I.}~\bibnamefont{Cholis}},
  \bibinfo{author}{\bibfnamefont{P.~C.} \bibnamefont{Breysse}},
  \bibnamefont{and}
  \bibinfo{author}{\bibfnamefont{M.}~\bibnamefont{Kamionkowski}},
  \bibinfo{journal}{Phys. Rev.} \textbf{\bibinfo{volume}{D95}},
  \bibinfo{pages}{103010} (\bibinfo{year}{2017}), \eprint{1611.01157}.

\bibitem[{\citenamefont{Mandic et~al.}(2016)\citenamefont{Mandic, Bird, and
  Cholis}}]{Mandic:2016lcn}
\bibinfo{author}{\bibfnamefont{V.}~\bibnamefont{Mandic}},
  \bibinfo{author}{\bibfnamefont{S.}~\bibnamefont{Bird}}, \bibnamefont{and}
  \bibinfo{author}{\bibfnamefont{I.}~\bibnamefont{Cholis}},
  \bibinfo{journal}{Phys. Rev. Lett.} \textbf{\bibinfo{volume}{117}},
  \bibinfo{pages}{201102} (\bibinfo{year}{2016}), \eprint{1608.06699}.

\bibitem[{\citenamefont{Clesse and
  García-Bellido}(2017{\natexlab{b}})}]{Clesse:2016ajp}
\bibinfo{author}{\bibfnamefont{S.}~\bibnamefont{Clesse}} \bibnamefont{and}
  \bibinfo{author}{\bibfnamefont{J.}~\bibnamefont{García-Bellido}},
  \bibinfo{journal}{Phys. Dark Univ.} \textbf{\bibinfo{volume}{18}},
  \bibinfo{pages}{105} (\bibinfo{year}{2017}{\natexlab{b}}),
  \eprint{1610.08479}.

\bibitem[{\citenamefont{Ananda et~al.}(2007)\citenamefont{Ananda, Clarkson, and
  Wands}}]{Ananda:2006af}
\bibinfo{author}{\bibfnamefont{K.~N.} \bibnamefont{Ananda}},
  \bibinfo{author}{\bibfnamefont{C.}~\bibnamefont{Clarkson}}, \bibnamefont{and}
  \bibinfo{author}{\bibfnamefont{D.}~\bibnamefont{Wands}},
  \bibinfo{journal}{Phys. Rev.} \textbf{\bibinfo{volume}{D75}},
  \bibinfo{pages}{123518} (\bibinfo{year}{2007}), \eprint{gr-qc/0612013}.

\bibitem[{\citenamefont{Baumann et~al.}(2007)\citenamefont{Baumann, Steinhardt,
  Takahashi, and Ichiki}}]{Baumann:2007zm}
\bibinfo{author}{\bibfnamefont{D.}~\bibnamefont{Baumann}},
  \bibinfo{author}{\bibfnamefont{P.~J.} \bibnamefont{Steinhardt}},
  \bibinfo{author}{\bibfnamefont{K.}~\bibnamefont{Takahashi}},
  \bibnamefont{and} \bibinfo{author}{\bibfnamefont{K.}~\bibnamefont{Ichiki}},
  \bibinfo{journal}{Phys. Rev.} \textbf{\bibinfo{volume}{D76}},
  \bibinfo{pages}{084019} (\bibinfo{year}{2007}), \eprint{hep-th/0703290}.

\bibitem[{\citenamefont{Garcia-Bellido
  et~al.}(2017)\citenamefont{Garcia-Bellido, Peloso, and
  Unal}}]{Garcia-Bellido:2017aan}
\bibinfo{author}{\bibfnamefont{J.}~\bibnamefont{Garcia-Bellido}},
  \bibinfo{author}{\bibfnamefont{M.}~\bibnamefont{Peloso}}, \bibnamefont{and}
  \bibinfo{author}{\bibfnamefont{C.}~\bibnamefont{Unal}},
  \bibinfo{journal}{JCAP} \textbf{\bibinfo{volume}{1709}}, \bibinfo{pages}{013}
  (\bibinfo{year}{2017}), \eprint{1707.02441}.

\bibitem[{\citenamefont{Bartolo
  et~al.}(2018{\natexlab{a}})\citenamefont{Bartolo, De~Luca, Franciolini,
  Peloso, and Riotto}}]{Bartolo:2018evs}
\bibinfo{author}{\bibfnamefont{N.}~\bibnamefont{Bartolo}},
  \bibinfo{author}{\bibfnamefont{V.}~\bibnamefont{De~Luca}},
  \bibinfo{author}{\bibfnamefont{G.}~\bibnamefont{Franciolini}},
  \bibinfo{author}{\bibfnamefont{M.}~\bibnamefont{Peloso}}, \bibnamefont{and}
  \bibinfo{author}{\bibfnamefont{A.}~\bibnamefont{Riotto}}
  (\bibinfo{year}{2018}{\natexlab{a}}), \eprint{1810.12218}.

\bibitem[{\citenamefont{Bartolo
  et~al.}(2018{\natexlab{b}})\citenamefont{Bartolo, De~Luca, Franciolini,
  Peloso, Racco, and Riotto}}]{Bartolo:2018rku}
\bibinfo{author}{\bibfnamefont{N.}~\bibnamefont{Bartolo}},
  \bibinfo{author}{\bibfnamefont{V.}~\bibnamefont{De~Luca}},
  \bibinfo{author}{\bibfnamefont{G.}~\bibnamefont{Franciolini}},
  \bibinfo{author}{\bibfnamefont{M.}~\bibnamefont{Peloso}},
  \bibinfo{author}{\bibfnamefont{D.}~\bibnamefont{Racco}}, \bibnamefont{and}
  \bibinfo{author}{\bibfnamefont{A.}~\bibnamefont{Riotto}}
  (\bibinfo{year}{2018}{\natexlab{b}}), \eprint{1810.12224}.

\bibitem[{\citenamefont{Cai et~al.}(2018)\citenamefont{Cai, Pi, and
  Sasaki}}]{Cai:2018dig}
\bibinfo{author}{\bibfnamefont{R.-g.} \bibnamefont{Cai}},
  \bibinfo{author}{\bibfnamefont{S.}~\bibnamefont{Pi}}, \bibnamefont{and}
  \bibinfo{author}{\bibfnamefont{M.}~\bibnamefont{Sasaki}}
  (\bibinfo{year}{2018}), \eprint{1810.11000}.

\bibitem[{\citenamefont{Bartolo et~al.}(2016)}]{Bartolo:2016ami}
\bibinfo{author}{\bibfnamefont{N.}~\bibnamefont{Bartolo}} \bibnamefont{et~al.},
  \bibinfo{journal}{JCAP} \textbf{\bibinfo{volume}{1612}}, \bibinfo{pages}{026}
  (\bibinfo{year}{2016}), \eprint{1610.06481}.

\bibitem[{\citenamefont{Bartolo
  et~al.}(2018{\natexlab{c}})\citenamefont{Bartolo, Domcke, Figueroa,
  García-Bellido, Peloso, Pieroni, Ricciardone, Sakellariadou, Sorbo, and
  Tasinato}}]{Bartolo:2018qqn}
\bibinfo{author}{\bibfnamefont{N.}~\bibnamefont{Bartolo}},
  \bibinfo{author}{\bibfnamefont{V.}~\bibnamefont{Domcke}},
  \bibinfo{author}{\bibfnamefont{D.~G.} \bibnamefont{Figueroa}},
  \bibinfo{author}{\bibfnamefont{J.}~\bibnamefont{García-Bellido}},
  \bibinfo{author}{\bibfnamefont{M.}~\bibnamefont{Peloso}},
  \bibinfo{author}{\bibfnamefont{M.}~\bibnamefont{Pieroni}},
  \bibinfo{author}{\bibfnamefont{A.}~\bibnamefont{Ricciardone}},
  \bibinfo{author}{\bibfnamefont{M.}~\bibnamefont{Sakellariadou}},
  \bibinfo{author}{\bibfnamefont{L.}~\bibnamefont{Sorbo}}, \bibnamefont{and}
  \bibinfo{author}{\bibfnamefont{G.}~\bibnamefont{Tasinato}}
  (\bibinfo{year}{2018}{\natexlab{c}}), \eprint{1806.02819}.

\end{thebibliography}

\end{document}